\documentclass[amsfonts,amsmath,prd,twocolumn,nofootinbib,a4paper,eqsecnum,superscriptaddress]{revtex4-2}
\newcommand{\beq}{ \begin{equation}}
\newcommand{\eeq}{\end{equation}}

\newcommand{\bsp}{\begin{split}}

\usepackage{epsfig,bbm,cancel}
\usepackage[normalem]{ulem}
\usepackage[colorlinks=true]{hyperref}
\usepackage{latexsym}
\usepackage[utf8]{inputenc}
\usepackage{amsmath}
\usepackage{amsfonts}
\usepackage{amssymb}
\usepackage{booktabs}
\usepackage{array}
\usepackage{tabularx}
\usepackage{multirow}
\usepackage{longtable}
\usepackage{ragged2e}
\usepackage{braket,booktabs}
\usepackage{braket,amsmath}
\usepackage{xcolor}
\usepackage{graphicx}
\usepackage{orcidlink} 

\begin{document}

\title{Effective theory of quantum black holes with Kretschmann scalar}

\author{G. I. B. Darman \orcidlink{0000-0003-0303-3628}}
\email{gema.ilham@alumni.ui.ac.id}
\affiliation{Departemen Fisika, FMIPA, Universitas Indonesia, Depok 16424, Indonesia}

\author{B. N. Jayawiguna \orcidlink{0000-0001-5920-8701}}
\email{nugrahabyon312@gmail.com}
\affiliation{Departemen Fisika, FMIPA, Universitas Indonesia, Depok 16424, Indonesia}

\author{M. F. Fauzi \orcidlink{0009-0005-3380-6005}}
\email{muhammad.fauzi@uclouvain.be}
\affiliation{Centre for Cosmology, Particle Physics and Phenomenology\textemdash CP3, Universit\'e Catholique de Louvain, Louvain-La-Neuve, B-1348, Belgium}
\affiliation{Departemen Fisika, FMIPA, Universitas Indonesia, Depok 16424, Indonesia}

\author{A. Sulaksono \orcidlink{0000-0002-1493-5013}}
\email{anto.sulaksono@sci.ui.ac.id}
\affiliation{Departemen Fisika, FMIPA, Universitas Indonesia, Depok 16424, Indonesia}

\date{\today}

\begin{abstract}
We investigate the thermodynamics, quasinormal modes (QNMs), and shadows of the effective theory of quantum black holes (BHs) by expanding the physical constants in terms of the Kretschmann scalar. The resulting effective metric can contain two horizons, which is typical for a quantum-corrected Schwarzschild spacetime. The second-order corrected metric only shifts the first-order corrected metric, allowing investigation of IR phenomena from the pure first-order corrected one. From a thermodynamic point of view, the metric leads to a stable cold remnant with zero entropy, compatible with the third law of thermodynamics. In the perturbation regime, we use massless scalar perturbations to extract the effective potential, QNMs, graybody factor, and total absorption cross section. We show that double-horizon BHs have a double light ring in the potential, and for a certain value $\Omega/\chi^2=0.01,$ the QNMs are similar to the GR case since the horizon is Schwarzschild-like. We also tested the correspondence between the QNMs and the graybody factor, and our results still hold with an error far below $1\%$. Our results also conclude that the higher the frequency, the bigger the cross-section area. For the optical observables, we observe that the opposite sign of $\Omega$ leads to distinct behavior of the BH's photon sphere and shadow radius. The Sagittarius A* shadow radius upper bound from the Event Horizon Telescope observation in the $2\sigma$ and $1\sigma$ significance set the upper limit of the quantum parameter $\Omega$ to $\Omega/\chi^2 \lesssim 0.3649$ and $\Omega/\chi^2 \lesssim 0.0105$, respectively.
\vspace{0.2 cm}

\end{abstract}

\maketitle
\thispagestyle{empty}
\setcounter{page}{1}

\section{Introduction}

Considerable effort has gone into modifying general relativity (GR) to incorporate quantum effects on BHs. However, a comprehensive theory of quantum gravity that naturally derives BH quantum effects remains an open problem. This opens the door to a model-independent approach to quantum gravity, either an effective theory or quantum-improved BHs from the renormalization group. 

Effective theories of quantum BHs are captured in~\cite{Binetti:2022xdi,del2024effective, cadoni2022effective}. In~\cite{Binetti:2022xdi}, the effective theory is constructed by expanding physical constants as a series of parameters related to proper distance. This approach aligns with the renormalized spacetime approach discussed in \cite{Bonanno_2000}. Proper distance can be used because it is a diffeomorphism-invariant quantity that does not change the algebra of the metric geometry. However, the expansion parameter of the physical constants is not limited to proper distance, since other diffeomorphism-invariant quantities could be explored \cite{Binetti:2022xdi, PhysRevD.104.066016, del2024effective}. 

Effective theory permits different alternatives as physical scale parameters, allowing us to use the scalar curvature (Kretschmann scalar) to reparameterize the metric. Kretschmann scalar has been used as a parameter for quantum improvements on BH in \cite{del2024effective,PhysRevD.104.066016}, however, it has not been utilized as a parameter for an effective theory that expands physical constants such as Planck constants and the Newton gravitational constant. In this article, the effective theory of BH with the Kretschmann scalar shows quantum features phenomenologically, which validate the theory.

In this article, we phenomenologically analyze the effective theory of quantum black holes using the Kretschmann scalar, including thermodynamics, QNMs, and shadows, to help justify the reparameterization adopted. Furthermore, we investigate higher-order improvements to the metric to assess the need to study second-order thermodynamics, QNMs, and shadows. A robust effective-theory model should also exhibit proper quantum features phenomenologically.

BH thermodynamics remains the cornerstone of BH testing. In this article, the fate of BH is a cold remnant with zero entropy, obeying the third law of thermodynamics \cite{Belgiorno:2002pm, socolovsky2023black}. In addition, the heat capacity behavior is consistent with other quantum black hole models \cite{Shi_2024, ali2015absence, ma2024thermodynamics}, where the heat capacity becomes positive at the remnant, indicating a thermodynamically stable remnant. Relevant thermodynamic quantities, i.e., Hawking radiation, local entropy, and heat capacity, match classical results for large BHs while exhibiting quantum features at the Planck scale. Hawking radiation also has two key components. The first is the blackbody spectrum associated with particle creation in the near-horizon region. The second is the grey-body factor, which encodes the frequency-dependent transmission probability for particles to penetrate the effective potential barrier and propagate to infinity. Determining the latter requires analyzing the associated scattering problem for a given potential and frequency. In the bound-state regime, the open nature of general relativity leads to a discrete set of complex frequencies, $\omega_{n}$, known as quasinormal modes (QNMs). The imaginary part of these frequencies implies a temporal decay of the wave amplitude, and such modes dominate the ringdown phase of gravitational wave (GW) signals. Because effective metric solutions are motivated by quantum corrections to classical spacetime, it is natural to investigate their semiclassical properties within a perturbative framework.

We also expand our discussion to the black hole's optical properties, focusing on the photon sphere and the resulting black hole shadow. The groundbreaking images captured by the Event Horizon Telescope (EHT) collaboration for the M87*~\cite{EventHorizonTelescope:2019dse,EventHorizonTelescope:2019uob,EventHorizonTelescope:2019jan,EventHorizonTelescope:2019ths,EventHorizonTelescope:2019pgp,EventHorizonTelescope:2019ggy,EventHorizonTelescope:2021bee,EventHorizonTelescope:2021srq,EventHorizonTelescope:2023gtd} and Sagittarius (Sgr) A*~\cite{EventHorizonTelescope:2022wkp,EventHorizonTelescope:2022apq,EventHorizonTelescope:2022wok,EventHorizonTelescope:2022exc,EventHorizonTelescope:2022urf,EventHorizonTelescope:2022xqj,EventHorizonTelescope:2024hpu,EventHorizonTelescope:2024rju} BHs have generated a huge interest in the observational signatures of BH shadows. By providing direct visual access to the strong-field regime, these observations offer a powerful new tool for testing fundamental physics. The obtained $1\sigma$ and $2\sigma$ observational bounds on the shadow radius~\cite{Vagnozzi:2022moj} have enabled researchers to constrain physics theories beyond general relativity~(see, e.g., Refs.~\cite{AlGhifari:2026vil, Bonanno:2025dry, Raza:2025ohk, Jha:2024eyb, Cao:2024vtq, Vachher:2024ait, Ali:2024ssf, Heidari:2024bkm, Chen:2024mlr, Konoplya:2024lch} for recent studies related to quantum-motivated corrections to BH geometry). Moreover, the upcoming Black Hole Explorer mission is expected to yield higher-resolution BH images~\cite{Johnson:2024ttr}, enabling us to probe fundamental physics by observing the BH shadow and photon rings with greater precision~\cite{Lupsasca:2024xhq}.

The paper is organized as follows. Section (\ref{Sec: Quantum Schwarzschild}) contains the newly corrected metric and its analysis. Sections (\ref{Sec 3: Thermodynamics})- \ref{Sec: shadow} present the phenomenological aspects, including thermodynamics, QNM, and shadow analyses. Section (\ref{Sec 5: Conclusions}) is devoted to the conclusion.  

\section{Quantum Schwarzschild Metric}\label{Sec: Quantum Schwarzschild}

The classical Schwarzschild metric reads 
\begin{equation}
ds^2=-f_0(r)dt^2+\frac{dr^2}{f_0(r)}+r^2(d\theta^2+\sin^2\theta d\phi^2),
\end{equation}
where $f_0(r)=1-\frac{2G_NM}{r}$, with $G_N$ as Newton's gravitational constant. In order to see deviation from the classical results, we expand around the coupling constant $G_NM_P^2$ with the Kretschmann scalar instead of the proper distance~\cite{Binetti:2022xdi, PhysRevD.104.066016}. The expansion of $G_NM_P^2$ can be done by scaling the mass and position parameters with Planck units as follows,
\begin{equation}
\chi:=\frac{M}{M_{\rm P}}, \qquad z:=\frac{r}{l_{\rm P}},
\end{equation}
where $M_P$ is the Planck mass, $l_P=M_P^{-1}$ is the Planck length. Thus, we can rewrite the lapse function as follows,
\begin{equation}
f_0(z) \rightarrow f(z)=1-\frac{2\chi}{z}G_{\rm N}M_{\rm P}^2,
\end{equation}
where $G_NM_P^2$ will be expanded with the Kretschmann scalar.

In~\cite{Binetti:2022xdi}, the coupling constant $G_NM_P^2$ is expanded as follows,
\begin{equation}\label{Eq: Binetti-Expansion}
G_{\rm N}M_{\rm P}^2=1+\Sigma^{\infty}_{n=1}\frac{\Omega_n}{d^{2n}_{n-1}(z)},
\end{equation}
where the coefficient $\Omega$ is some coupling constant and $d(z)$ is the proper distance. However, in this work, we investigate the Kretschmann scalar as an alternative. In \cite{PhysRevD.104.066016}, the scale-identification $k$ is related to the Kretschmann scalar as follows, 
\begin{equation}
k^4(z) = \xi^4 K(z),
\end{equation}
where $\xi$ is some constant and $K(z)$ is the Kretchsmann scalar. Thus, following Eq.~(\ref{Eq: Binetti-Expansion}), the coupling constants could be expanded as follows,
\begin{equation}
G_{\rm N}M_{\rm P}^2=1+\Sigma_{n=1}^\infty \Omega_nK_{n-1}^{n/2}(z),
\end{equation}
where the scale-identification reads $k(z) = \frac{\xi}{d(z)}$ \cite{Bonanno_2000}.

The Kretschmann scalar for the classical Schwarzchild solution reads
\begin{equation}
K_0=\frac{48G_{\rm N}^2M^2}{r^6}=\frac{48G_{\rm N}^2 M_{\rm P}^2 \chi^2}{z^6l_{\rm P}^6},
\end{equation} 
where for minimum quantum deformation and convenience in investigating the observables, we can take $G_N=M_P=l_P=1$ just for the Kretschmann scalar, resulting
\begin{equation}
K_0(z)=\frac{48\chi^2}{z^6}.
\end{equation}
Thus, the first-order quantum corrected lapse function $f_1(z)$ reads
\begin{equation}\label{Eq: 1st order lapse}
f_1(z)=1-\frac{2\chi}{z}\left[1+\Omega_1\sqrt{K_0(z)}\right]=1-\frac{2\chi}{z}\left(1+\frac{4\sqrt{3}\chi\Omega_1}{z^3}\right),
\end{equation}
where it returns to the classical form at $\Omega=0$.

\begin{figure*}[t!]
	\centering
	\includegraphics[width=0.49\linewidth]{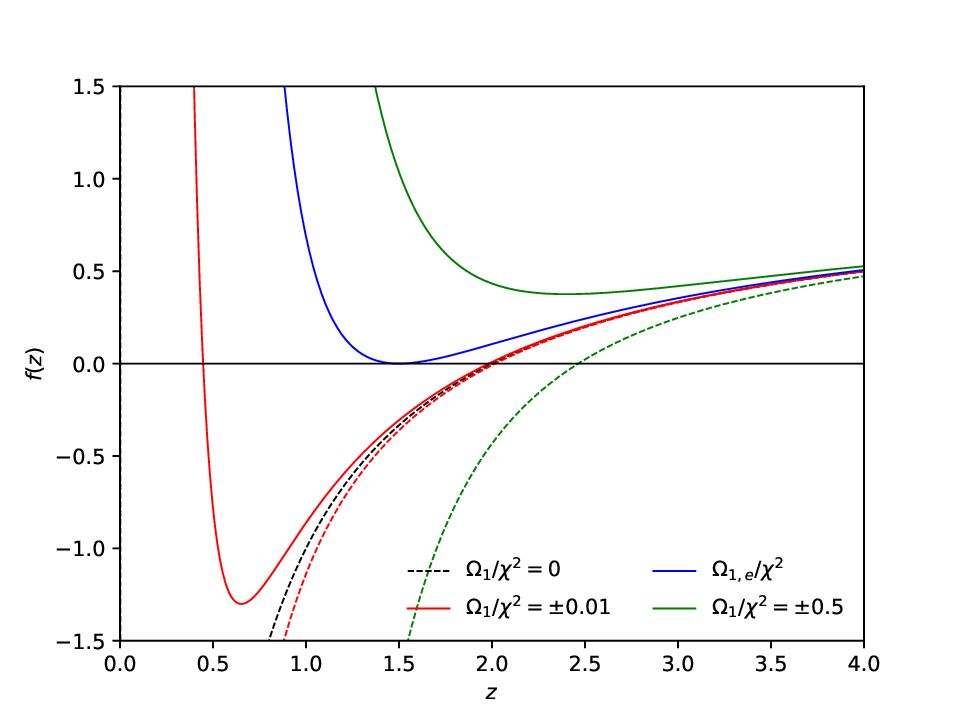}
	\includegraphics[width=0.49\linewidth]{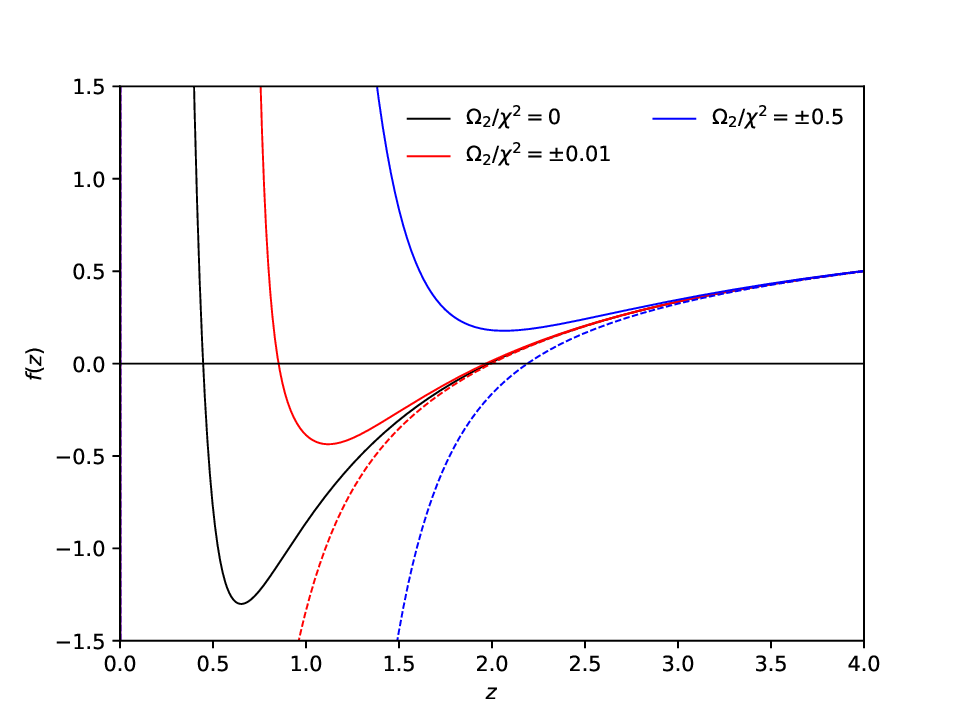}
	\caption{ 
		(Left) First-order lapse function $f_1(z)$ from
		Eq.~\eqref{Eq: 1st order lapse}, with various $\Omega_1/\chi^2$. The dashed lines indicates the positive part of $\Omega_1/\chi^2$.
		The solid blue curve corresponds to the extremal configuration,
		$\Omega_{1,e}=-9\sqrt{3}/128$, given in
		Eq.~\eqref{Eq: extremal-Schwarzschild}.
		(Right) Second-order lapse function $f_2(z)$ from
		Eq.~\eqref{Eq: Second Order}, with $\Omega_1/\chi^2=-0.01$ and various $\Omega_2/\chi^2$. The dashed lines indicates the positive part of $\Omega_2/\chi^2$.}
	\label{fig:fz_Schwarz}
\end{figure*}

The implication of Eq.~\eqref{Eq: 1st order lapse} is that it can have up to two horizons, i.e., two real solutions to $f(z_{\pm})=0$. The plot in Fig.~(\ref{fig:fz_Schwarz}) shows that two horizons appear for $\Omega <0$. As we increase the $\Omega$ parameter, both inner and outer horizons approach each other until they coincide at one extremal value. The extremal case can be obtained by the following conditions, which are $f(z_e)=0$ and $f'(z_{e})=0$, and the result is given by

\begin{eqnarray}\label{Eq: extremal-Schwarzschild}
z_{\rm e} = \frac{3}{2},~~~\textrm{and}~~~\Omega_{1,\rm e} = -\frac{9\sqrt{3}}{128},
\end{eqnarray}

where the subscript ``e" denotes the quantity at the extremal condition at a certain value of scaled mass $\chi$ ($\chi=1$). The latter results show that the extremal case occurs only for $\Omega<0$. The parameter $\Omega<\Omega_{1,e}$ results in naked singularity. For $\Omega>0$, the horizon is Schwarzschild-like and shifts as $\Omega$ increases.

If the phenomenological aspects of the first-order corrected metric have a unique signature, the need to analyze the higher-order metric diminishes significantly. However, it would be interesting to investigate the behavior of at least the UV-regime of the second-order metric. The first-order derivative of the Kretschmann scalar for the second-order corrected metric reads

\begin{equation}
K_1(z)=\frac{48 \chi^2 \left(1872 \chi^2 \Omega ^2+40 \sqrt{3} \chi z^3 \Omega +z^6\right)}{z^{12}},
\end{equation}
allowing the second order metric to be 
\begin{widetext}
	\begin{equation}\label{Eq: Second Order}
	f_2(z) = 1 - \frac{2\chi}{z}
	\left[
	1 + \frac{4\sqrt{3}\chi\Omega_1}{z^3}
	+ \frac{48\chi^2\Omega_2}{z^{12}}
	\left(
	1872\chi^2\Omega_1^2
	+ 40\sqrt{3}\chi\Omega_1 z^3
	+ z^6
	\right)
	\right],
	\end{equation}
\end{widetext}
where it returns to the classical solution for $\Omega_1=\Omega_2=0$, and the first-order solution for $\Omega_2=0$ and $\Omega_1 \neq 0$. 

Fig.~(\ref{fig:fz_Schwarz}) displays the lapse functions for $\Omega_2<0$ with fixed $\Omega_1=-0.01$. This value of $\Omega_1$ was chosen because Fig.~(\ref{fig:fz_Schwarz}) shows two horizons, which may reflect a quantum aspect. Meanwhile, the case $\Omega_2>0$ only shifts the classical function. This means the behavior of the first-order solution resembles that of the second-order one, allowing us to focus only on the first-order IR phenomena.

In the next sections, we will only use Eq.~(\ref{Eq: 1st order lapse}) to analyze the effects of quantum corrections on thermodynamics, QNMs, and shadows. We will investigate the impact for various possible values of the constant $\Omega_1$. Later, we will set $\Omega_{1} \equiv \Omega$.

\section{Thermodynamics}
\label{Sec 3: Thermodynamics}

The Hawking temperature of the corresponding quantum-corrected Schwarzcschild reads
\begin{equation}
\label{Eq: Hawking temperature}
T_{\rm H}(z_+)=\frac{\kappa}{2\pi}=\frac{f_1'(z_+)}{4\pi}=\frac{\chi}{2\pi z_+^2}\left(1+\frac{16\sqrt{3}\Omega\chi}{z_+^3}\right),
\end{equation}
which can be obtained from surface gravity $\kappa$ with $G_N=M_P=l_P=1$ for minimum deformation.





\begin{figure*}[t!]
	\centering
	\includegraphics[width=0.49\textwidth]{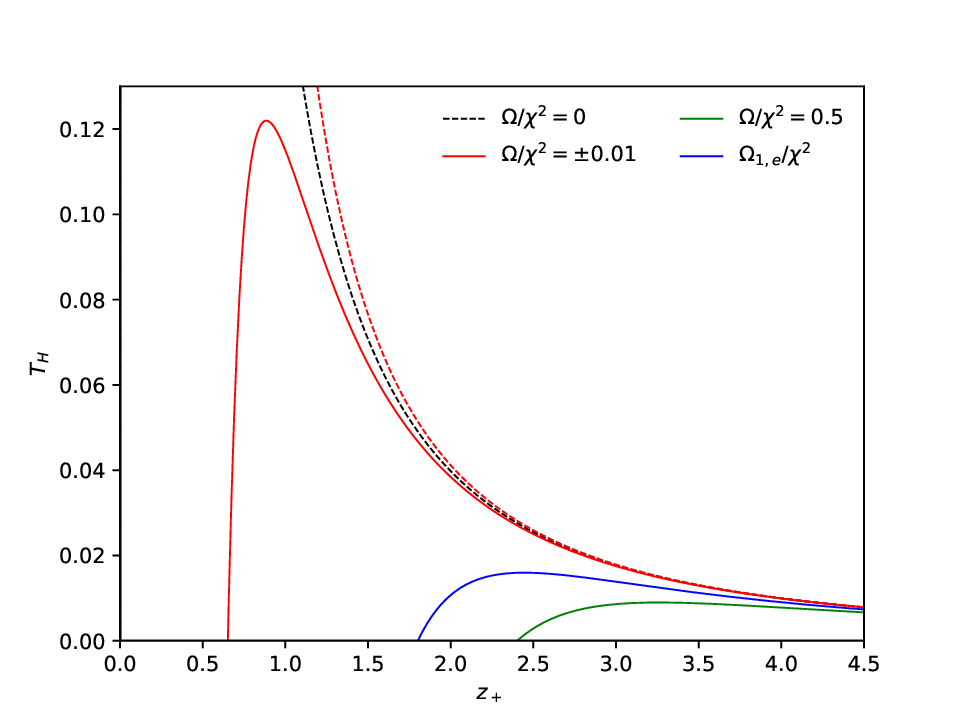}
	\includegraphics[width=0.49\textwidth]{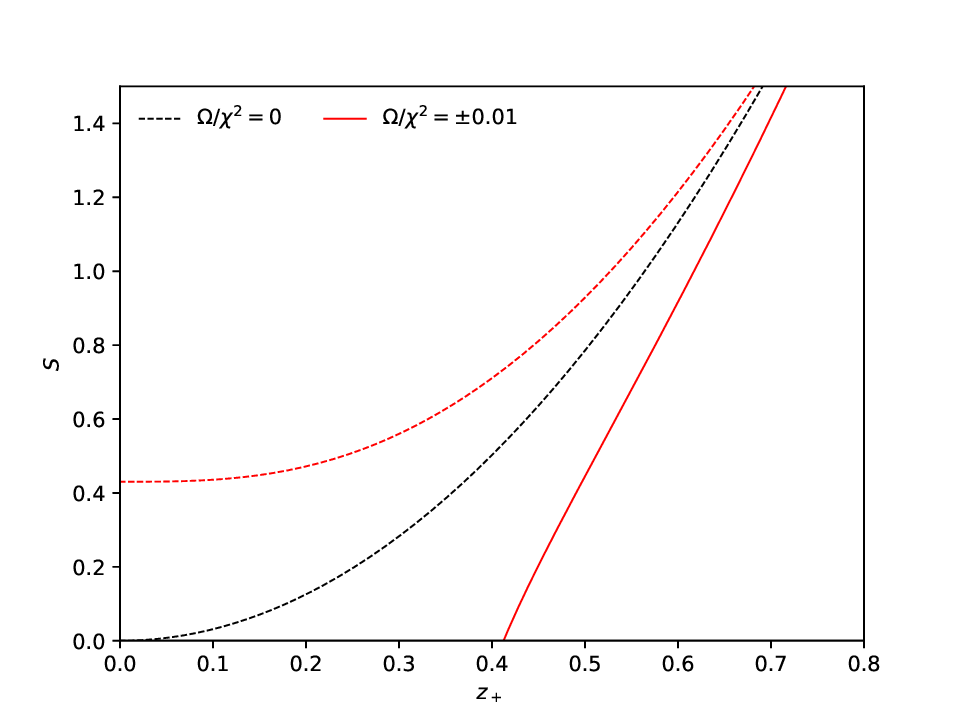}
	\includegraphics[width=0.49\textwidth]{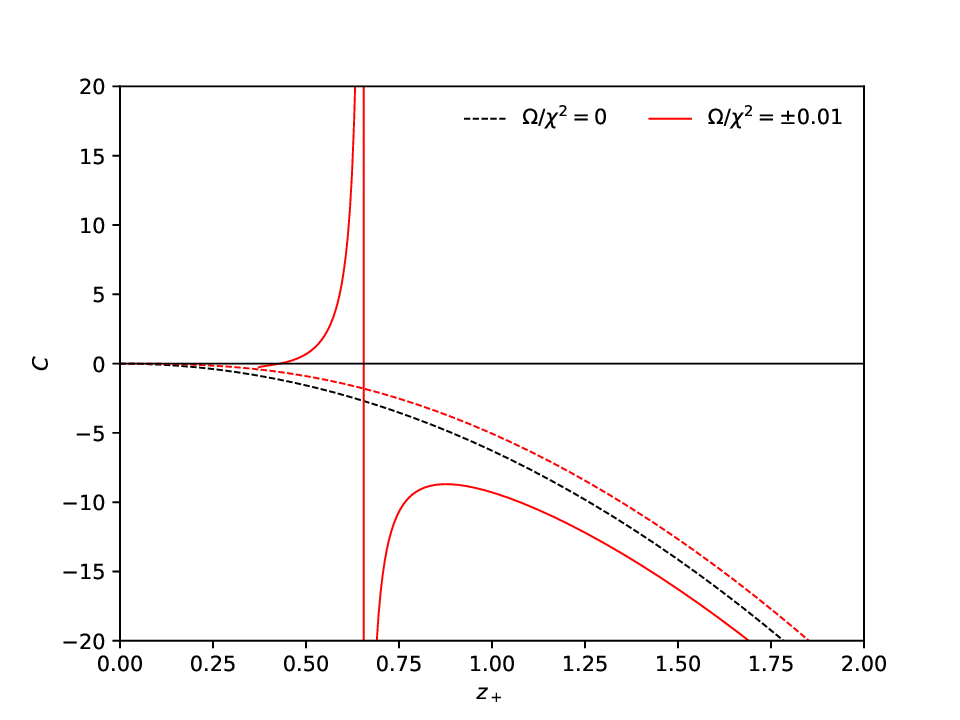}
	\caption{
		(Top left) Hawking temperature $T_H(z_+)$ from
		Eq.~\eqref{Eq: Hawking temperature}, (Top right) Entropy $S(z_+)$ from Eq.~\eqref{Eq: entropy}, and (Bottom) Heat capacity $C(z_+)$ from Eq.~\eqref{Eq: Heat capacity}, with various $\Omega/\chi^2$.}
	\label{fig:TH_Schwarz}
\end{figure*}

The plot in Fig.~(\ref{fig:TH_Schwarz}) for $\Omega<0$ leads to the presence of a cold remnant. However, for $\Omega \geq 0$, the Hawking temperature becomes singular. This supports the finding in Fig.~(\ref{fig:fz_Schwarz}) that the case $\Omega>0$ resembles the classical solution ($\Omega=0$).

Thermodynamic analysis usually involves investigating heat capacity and local entropy to study possible phase transitions and other features of quantum corrections. Regions where the heat capacity becomes zero or positive indicate a phase transition from a thermodynamically unstable to a stable BH. The entropy is obtained as follows,
\begin{equation}
S = \int \frac{1}{T_{\rm H}} \left(\frac{\partial\chi}{\partial z_+}\right)_{\Omega} dz_+,
\end{equation}
where the scaled mass $\chi(z_+)$ is obtained from $f(z_+)=0$. The entropy reads
\begin{equation}\label{Eq: entropy}
S = \pi  z_+ \sqrt{8 \sqrt{3}~ \Omega +z^2_+}-8\pi~ \Omega\sqrt{3}  \log \left(\sqrt{8 \sqrt{3} \Omega +z^2_+}+z_+\right),
\end{equation}
in which, it returns to the classical result at $\Omega=0$ ($S_{classical}=\pi z_+^2$). 
Fig.~(\ref{fig:TH_Schwarz}) implies that a remnant is the end state of quantum black holes when $\Omega<0$. This supports the finding in Fig.~(\ref{fig:TH_Schwarz}) since, for $\Omega<0$, the remnant's Hawking temperature is also zero. The entropy behavior in the case of $\Omega>0$ is in tension with the third law of thermodynamics—at the end state, a BH should have zero entropy and Hawking temperature. In general, the entropy behavior for $\Omega<0$ is also consistent with its corresponding metric and Hawking temperature.

On the other hand, the heat capacity is obtained as follows,
\begin{equation}
C = T_{\rm H} \left( \frac{\partial S/\partial z_{+}}{\partial T_{\rm H}/\partial z_{+}}  \right)_{\Omega}
\end{equation}
where the result reads
\begin{widetext}
	\begin{equation}
	\label{Eq: Heat capacity}
	C = -\frac{2 \pi  z^3_+ \left(16 \sqrt{3} \Omega +z_+ \sqrt{8 \sqrt{3} \Omega
			+z^2_+}+z^2_+\right)}{z^3_++z^2_+ \sqrt{8 \sqrt{3} \Omega +z^2_+}+16 \sqrt{3} \Omega 
		\sqrt{8 \sqrt{3} \Omega +z^2_+}+32 \sqrt{3} \Omega  z_+},
	\end{equation}
\end{widetext}
where it also returns to the classical result of the heat capacity at $\Omega=0$ ($C_{classical}=-2\pi z_+^2$).
In Fig.~(\ref{fig:TH_Schwarz}), the stable BH case with $C>0$ emerges when $\Omega=0$. However, for $\Omega\geq 0$, no phase transition occurs since $C<0$ everywhere. In conclusion, the BH model in this article shows quantum behavior, thermodynamically, at $\Omega<0$. 

\section{Perturbation Field}

\begin{figure*}[ht!]
	\centering
	\includegraphics[width=0.45\textwidth]{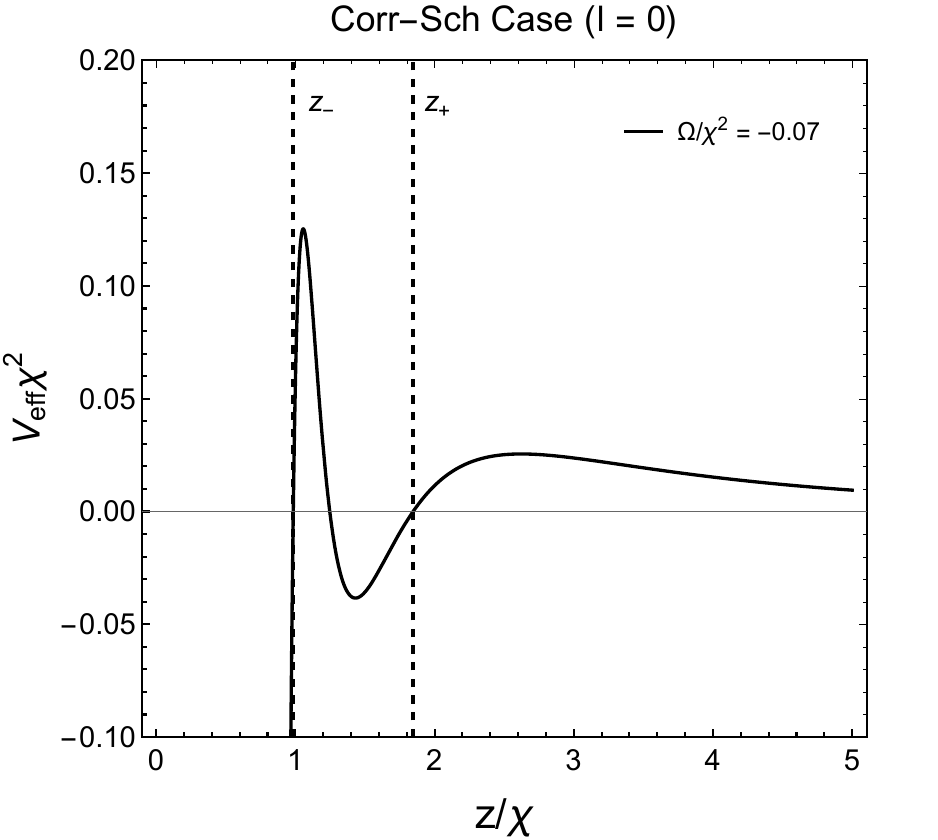}
	\includegraphics[width=0.45\textwidth]{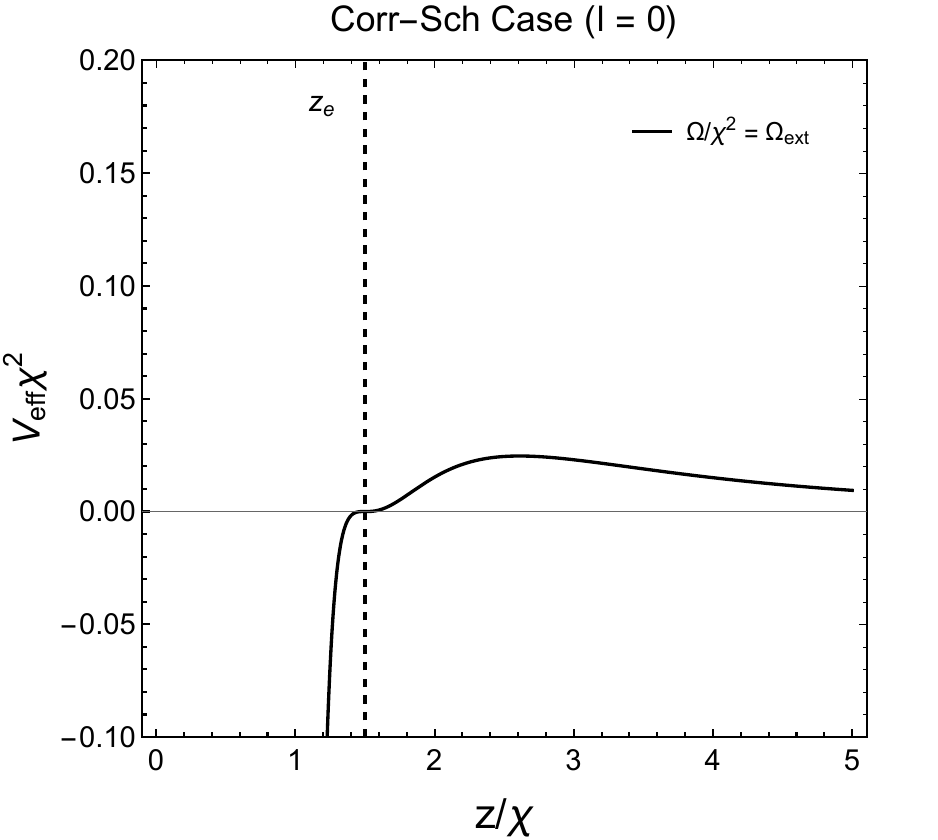}
	\includegraphics[width=0.45\textwidth]{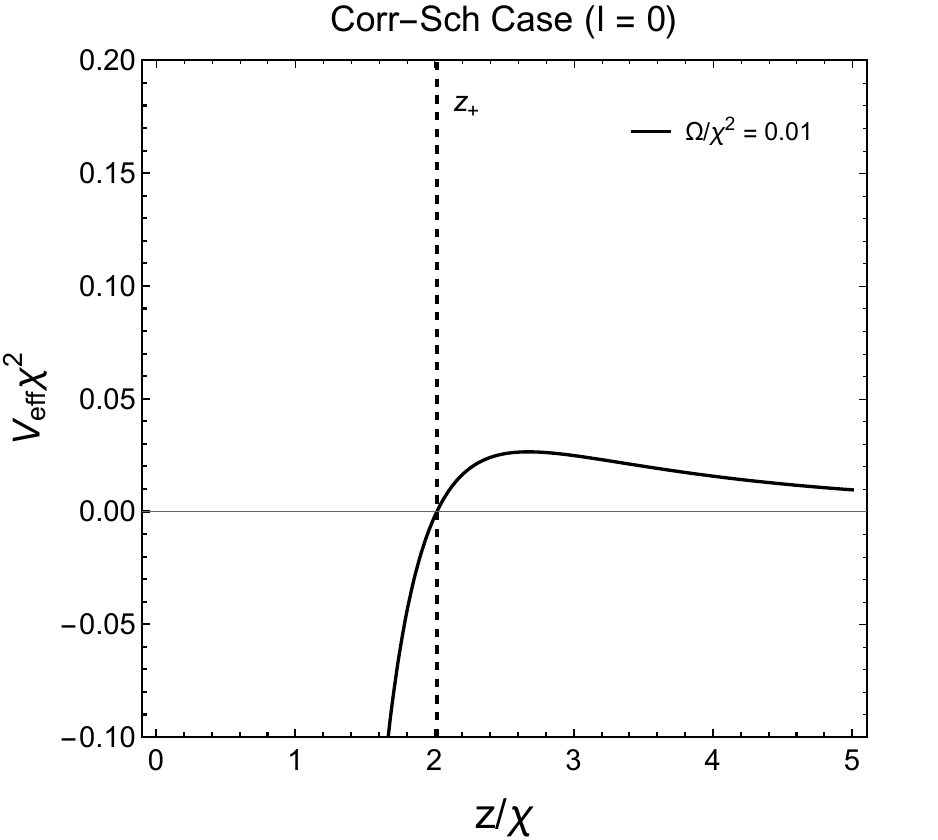}
	\caption{Effective potential versus the radial coordinate for $l=0$ with $\Omega/\chi^2=-0.07$, $\Omega/\chi^2=\Omega_{e}$, and one representative value for the positive branch, $\Omega/\chi^2=0.01$.}
	\label{fig:veff}
\end{figure*}

We discussed black hole configurations in the Kretchmann scalar expansion series and examined their phenomenological aspects, including the shadow and the intensity around the object. On the other hand, we can relate the characteristic wave frequency and the transmission rate through the grey-body factor and the absorption cross section to possible observations. The latter can be done by studying the black hole perturbation, and in this work, we used the simple massless scalar wave, which can be written as
\begin{equation}
\nabla_{\mu}\nabla^{\mu}\Psi=0.
\end{equation}
The contravariant derivative acting on the scalar field can be solely transformed into the ordinary derivative, while the covariant derivative on the far left can be evaluated into 
\begin{equation}
\frac{1}{\sqrt{-g}}\partial_{\mu}(\sqrt{-g}g^{\mu\nu}\partial_{\nu}\Psi)=0,
\end{equation}
where $\Psi$ depends on the space and time components, plugging in the metric $g^{\mu\nu}$, raising the indices, and using the separable equation, we obtain a simplified form in Tortoise coordinates.
\begin{equation}
\left[ \frac{\partial^2}{\partial r_{*}^2} +\omega^2- V \right]\psi=0,
\end{equation}
where the coordinate definition and the effective potential can be defined as
\begin{eqnarray}
\frac{dr}{dr_{*}} = \frac{1}{f(r)},~~~\textrm{and}~~~V(r) = f(r) \left( \frac{l(l+1)}{r^2}  + \frac{f'(r)}{r} \right),\nonumber\\
\end{eqnarray}
The constant $l$ is the eigenvalue of the angular equation that determines the BH barrier potential. The properties of the Schwarzschild first correction for scalar waves are encoded in the effective potential, as shown in Fig.~\ref{fig:veff}. The plot shows the dimensionless parameter as a function of the radial coordinate $z$ with $l=0$. The profile can be understood by considering both the negative branch ($\Omega<0$) and the positive branch ($\Omega>0$). The metric in the positive branch has one horizon; therefore, the effective potential has only Schwarzschild-like properties. The interesting part comes from the negative branch, where dynamical horizons appear. When $\Omega/\chi^2 = -0.07$, two light rings exist. However, the physically observed light ring is the outer one. As the negative parameter increases, the inner and outer horizons approach each other until they merge at the extreme case ($\Omega/\chi^2 = \Omega_{e}$), where they coincide at $z_e$; the inner light ring disappears when the BH surface merges.

\begin{table}[h]
	\centering
	\setlength{\tabcolsep}{10pt}
	\begin{tabular}{ccc}
		\hline\hline
		\multicolumn{3}{c}{Corr-Sch Case ($l=0$)} \\
		\hline
		$\Omega/\chi^2$ & $\omega_r$ & $|\omega_i|$ \\
		\hline
		$ 0$~~ & $1.1 \times10^{-1}$ & $1.01 \times10^{-1}$ \\
		$ -0.07$~~ & $1.23 \times10^{-1}$ & $8.26 \times10^{-2}$ \\
		$ \Omega_{e}$ & $1.13 \times10^{-1}$ & $7.52 \times10^{-2}$ \\
		$ 0.01$  & $1.1 \times10^{-1}$ & $1.01 \times10^{-1}$ \\
		\hline\hline
	\end{tabular}
	\caption{The quasinormal mode of the first order of Schwazrschild black hole with different case of $\Omega$.}
	\label{tab:1}
\end{table}

The quasinormal mode (QNM) can thus be explored through $\omega=\omega_{r}+i\omega_{I}$, where $ \omega _ {r}$ and $ \omega _ {I} $ are the real and imaginary parts of the frequency. In this terminology, the wave frequency dissipates through space and slowly goes to zero at large distances, and its stability depends on the sign of the damping time, $\omega_{I}$. The $\omega_{I}>0 ~(<0)$ is stable (unstable). To evaluate this, one needs the proper boundary condition, and we consider the case where the wave is purely outgoing. The method could be numerical brute force, but instead, we use a good approximation of the WKB formalism. With the improvement in the general case, the formal equation can be written as 
\begin{equation}
\label{k}
\mathcal{K}\equiv \frac{iQ_{\omega0}}{\sqrt{2 Q_{\omega0}''}}- \sum^{N}_{i=2} \Lambda_{i} = n+ \frac{1}{2},
\end{equation}
where $Q_{\omega0}\equiv \omega^2 - V_{0}$ and $\mathcal{K}$ is the quantity evaluated at the peak potential and the variable we will use in the next properties, respectively; the correction $\Lambda_{i}$ is the continued version of the expansion depending on how we set the upper boundary $N$. In the $N=6$ case, the correction is given in \cite{Konoplya2004}. Lastly, the variable $n$ is called the overtone number. According to \cite{Konoplya:2019hlu}, the smallest error occurs at the $ 6$th and $ 7$th orders; thus, we follow them by using the Wolfram Mathematica Packages \cite{Konoplya2004}. 

To calculate the quasinormal modes, one needs information about the potential peak. In this case, we choose the potential peak outside the outer horizon. The values are presented in Table~\ref {tab:1}. The imaginary part is negative, indicating unstable configurations. When $\Omega/\chi^2=0.01$, we obtain a result similar to the GR case (after rounding up), since the metric is Schwarzschild-like.

\subsection{Greybody factor}

\begin{figure*}[t!]
	\centering
	\includegraphics[width=0.47\linewidth]{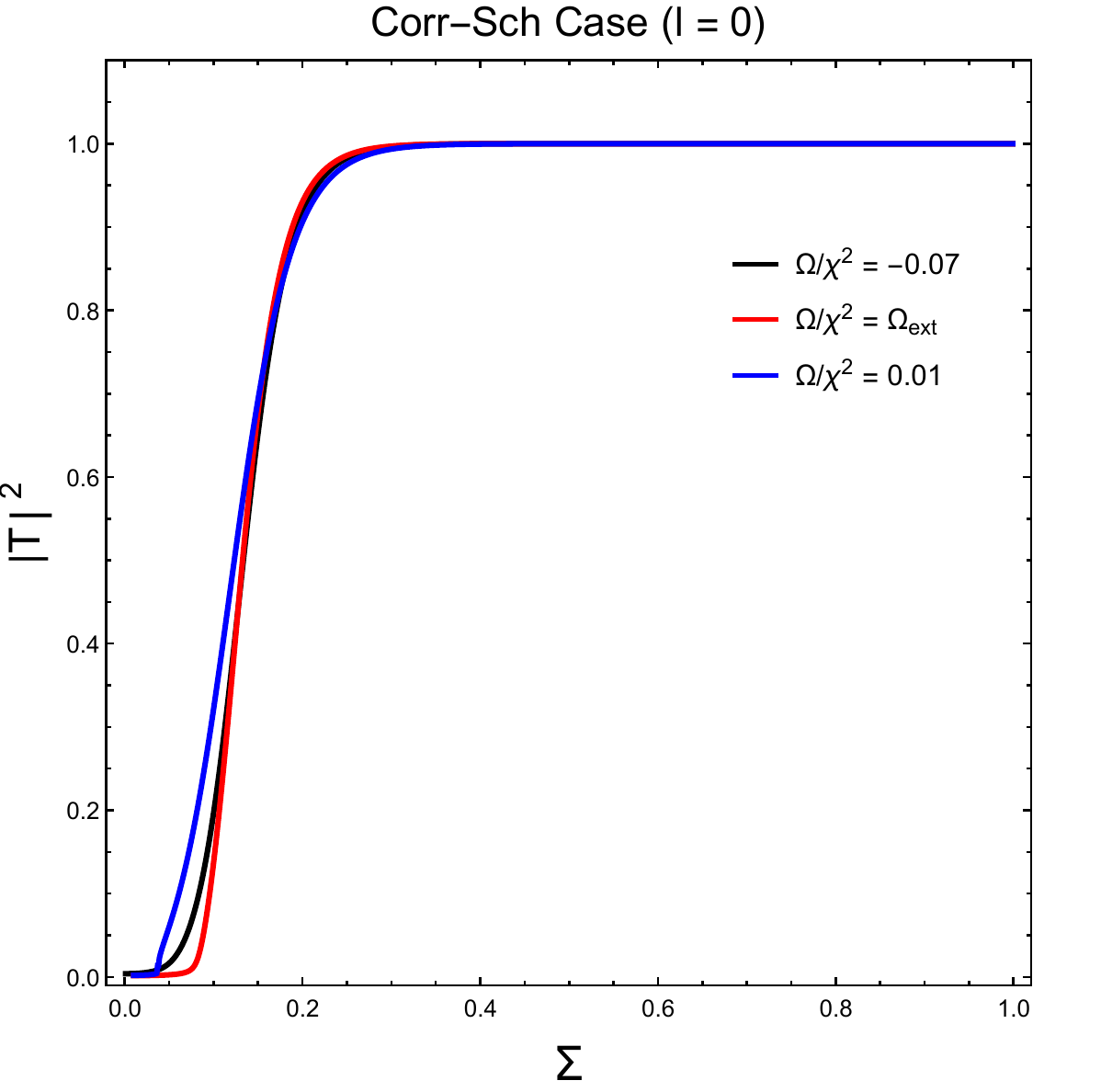}
	\includegraphics[width=0.47\linewidth]{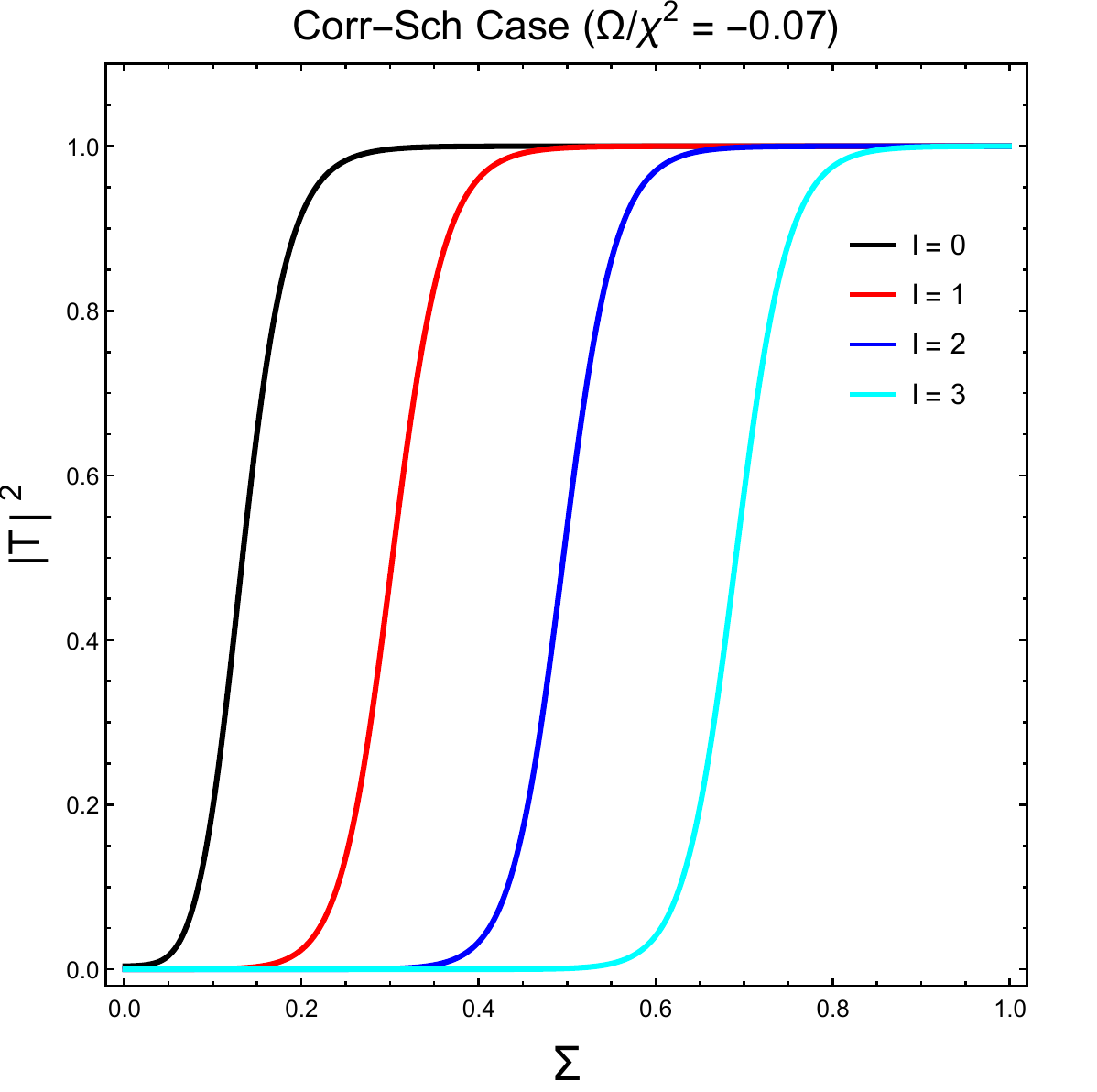}
	\caption{The transmission coefficient (greybody) in terms of continuous frequency, $\Sigma$, with various $\Omega/\chi^2$ (left) and various $l$ (right). All the waves have bigger probabilities to travel with a very high frequency.}
	\label{fig:T2}
\end{figure*}

The spacetime metrics discussed in the previous section are inspired by quantum properties in classical GR, allowing a deviation from the Schwarzschild and related solutions. Therefore, it is intriguing to analyze other properties within the same semiclassical framework. In the semiclassical picture, the black hole radiates via outgoing waves with a thermal spectrum at the surface gravity evaluated at the horizon. To an observer at a large distance, the wave is not perfect blackbody radiation due to the backreaction of spacetime curvature outside the horizon. This yields grey-body factors with frequency-dependent transmission coefficients that modulate the emission spectrum. In the mathematical context, the greybody factor calculates the transmission probability in the scattering problem.
\begin{equation}
\Gamma(\Sigma) = |\mathcal{T}|^2 = \frac{1}{1+e^{2\pi i \mathcal{K}(\Sigma)}},
\end{equation}
where $\mathcal{K}$ is defined in \eqref{k} with frequency ($\Sigma$)-dependent. Note that $\omega$ is a discrete set of quasinormal modes, and $\Sigma$ is the continuous frequency in the scattering state. Fig.~\ref{fig:T2} shows the greybody factors as a function of frequency. When we vary the quantum correction with a fixed $l$, the profiles remain largely unchanged. The profiles shift when we consider a higher potential barrier, since it affects the continuous frequency range. The higher the barrier, the higher the frequency must be to allow transmission.

\begin{figure*}[t!]
	\centering
	\includegraphics[width=0.46\linewidth]{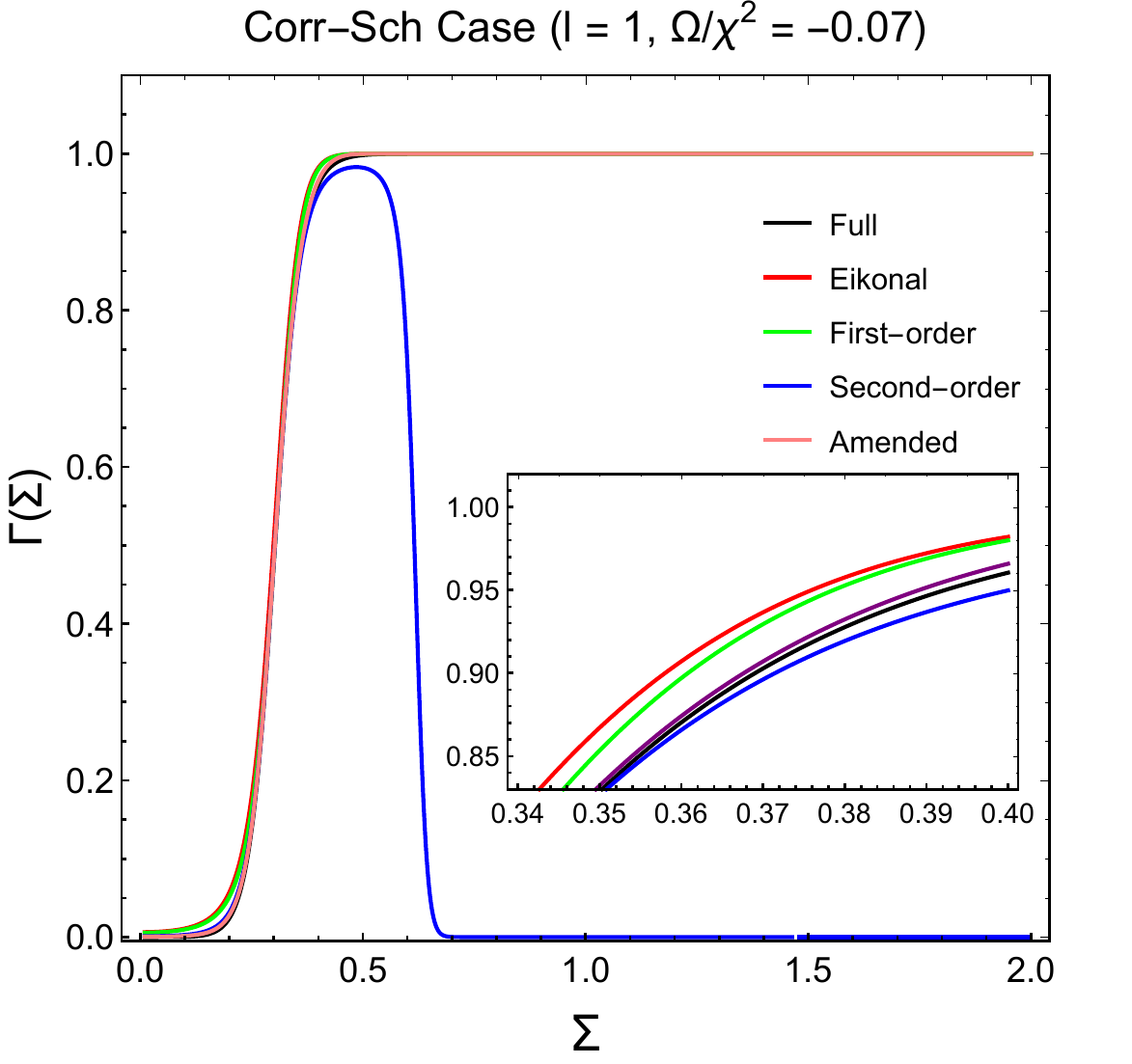}
	\includegraphics[width=0.46\linewidth]{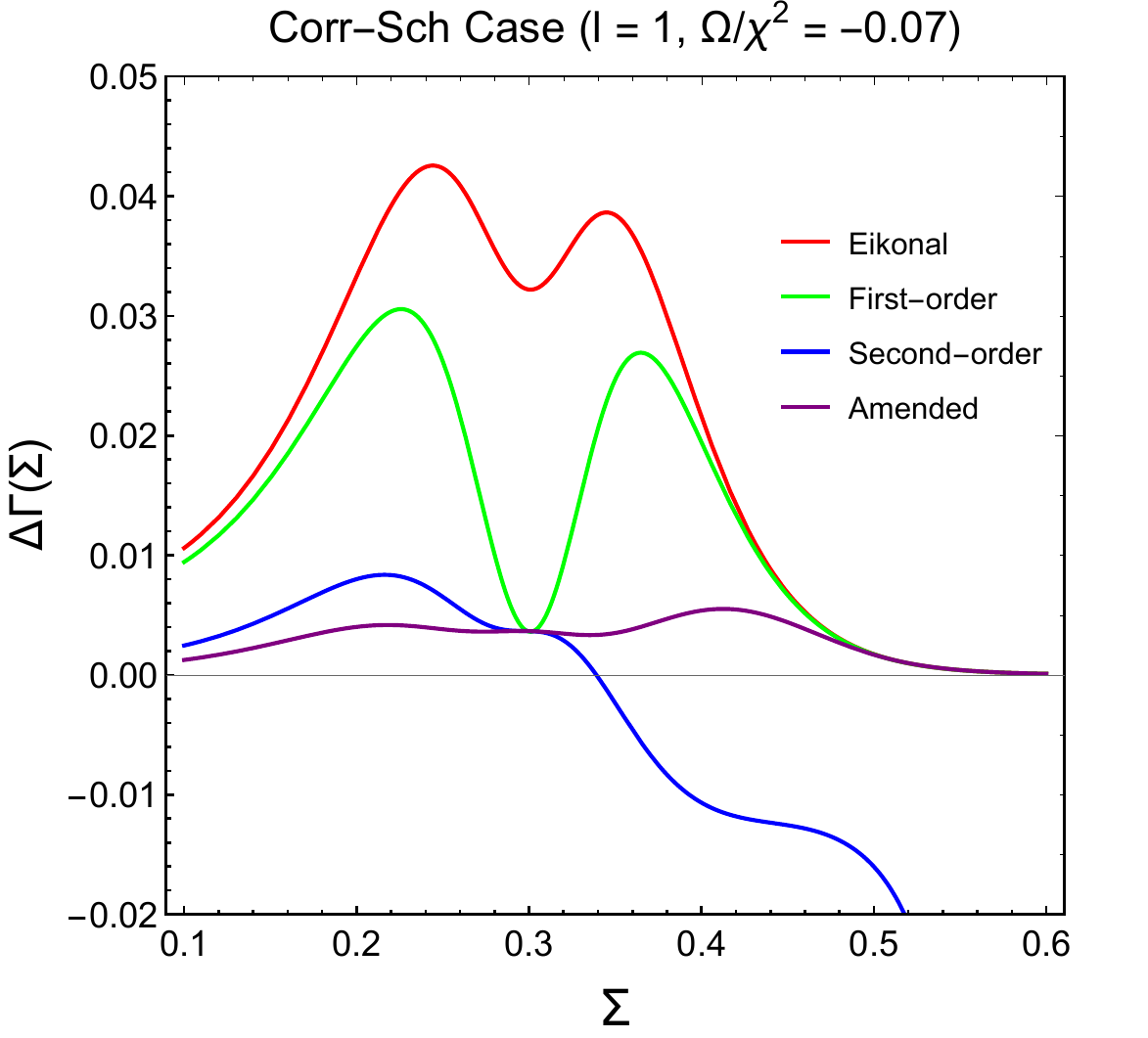}
	\caption{The greybody properties constructed by the quasinormal mode value (left), and the difference between the full greybody using WKB and the eikonal series (right).}
	\label{fig:corrected}
\end{figure*}

In the reference \cite{Konoplya:2024lir}, an alternative property of the scattering problem is the correspondence with the bound-state QNM through the eikonal limit: the expansion of $l$ in the WKB formula. Therefore, we can relate the $\omega$ and $\Sigma$ frequencies directly. The equation is given by

\begin{equation}
-i\mathcal{K} = -\frac{\Sigma^2-\Re(\omega_0)^2}{4\Re(\omega_0)\Im(\omega_0)} + \Delta_1 + \Delta_2 + \Delta_\textrm{f} + \mathcal{O}(l^{-3}),
\end{equation}
where the $\omega_{0}$ and $\omega_1$ are the fundamental mode ($n=0$) and 1st overtone mode ($n=1$), respectively. The $\Delta$ terms are given by
\begin{widetext}
	\begin{eqnarray}
	\Delta_1 &=& \frac{\Re(\omega_0)-\Re(\omega_1)}{16~\Im(\omega_0)} , \\ \Delta_2 &=& - \frac{\Sigma^2-\Re(\omega_0)^2}{32~\Re(\omega_{0})\Im(\omega_0)} \Bigg\lbrace \frac{[\Re(\omega_0)-\Re(\omega_1)]^2}{4~\Im(\omega_0)^2}-\frac{3~\Im(\omega_0)-\Im(\omega_1)}{3\Im(\omega_0)}     \Bigg\rbrace  + \frac{[\Sigma^2-\Re(\omega_0)^2]^2}{16~\Re(\omega_{0})^3\Im(\omega_0)} \Bigg \lbrace 1+ \frac{\Re(\omega_0)[\Re(\omega_0)-\Re(\omega_1)]}{4~\Im(\omega_0)^2}   \Bigg \rbrace , \nonumber \\ \\ \Delta_\textrm{f} &=& -\frac{[\Sigma^2-\Re(\omega_0)^2]^3}{32~\Re(\omega_0)^5 \Im(\omega_{0})} \Bigg( 1+ \frac{\Re(\omega_0)[\Re(\omega_0)-\Re(\omega_1)]}{4~\Im(\omega_0)^2}  + \Re(\omega_0)^2 \Bigg\lbrace \frac{[\Re(\omega_0)-\Re(\omega_1)]^2}{16~\Im(\omega_{0})^4} -\frac{3\Im(\omega_0)-\Im(\omega_1)}{12~\Im(\omega_0)} \Bigg\rbrace  \Bigg).
	\end{eqnarray}
\end{widetext}
We restrict the expansion to the sixth-order WKB formula and the $\mathcal{O}(l^{-3})$ expansion. The latter expression allows us to approximate the grey body factors using the first two overtone numbers, namely $\omega_0$ and $\omega_1$, from the QNM. Later, we show the difference using the formula $\Delta\Gamma(\Sigma)\equiv|\Gamma_{\textrm{WKB}}-\Gamma_{\textrm{eikonal}}|$. In Fig.~\ref{fig:corrected}, we compare the greybody data profile from a full numerical simulation with the eikonal limit using the quasinormal mode value. In the left panel, the higher-order eikonal is very close to the numerical result. The $\Delta\Gamma$ profile plot further supports this. The blue one is the smallest among the two, with the difference below $1\%$.

\subsection{Absorption cross section}

\begin{figure}[t!]
	\centering
	\includegraphics[width=0.8\linewidth]{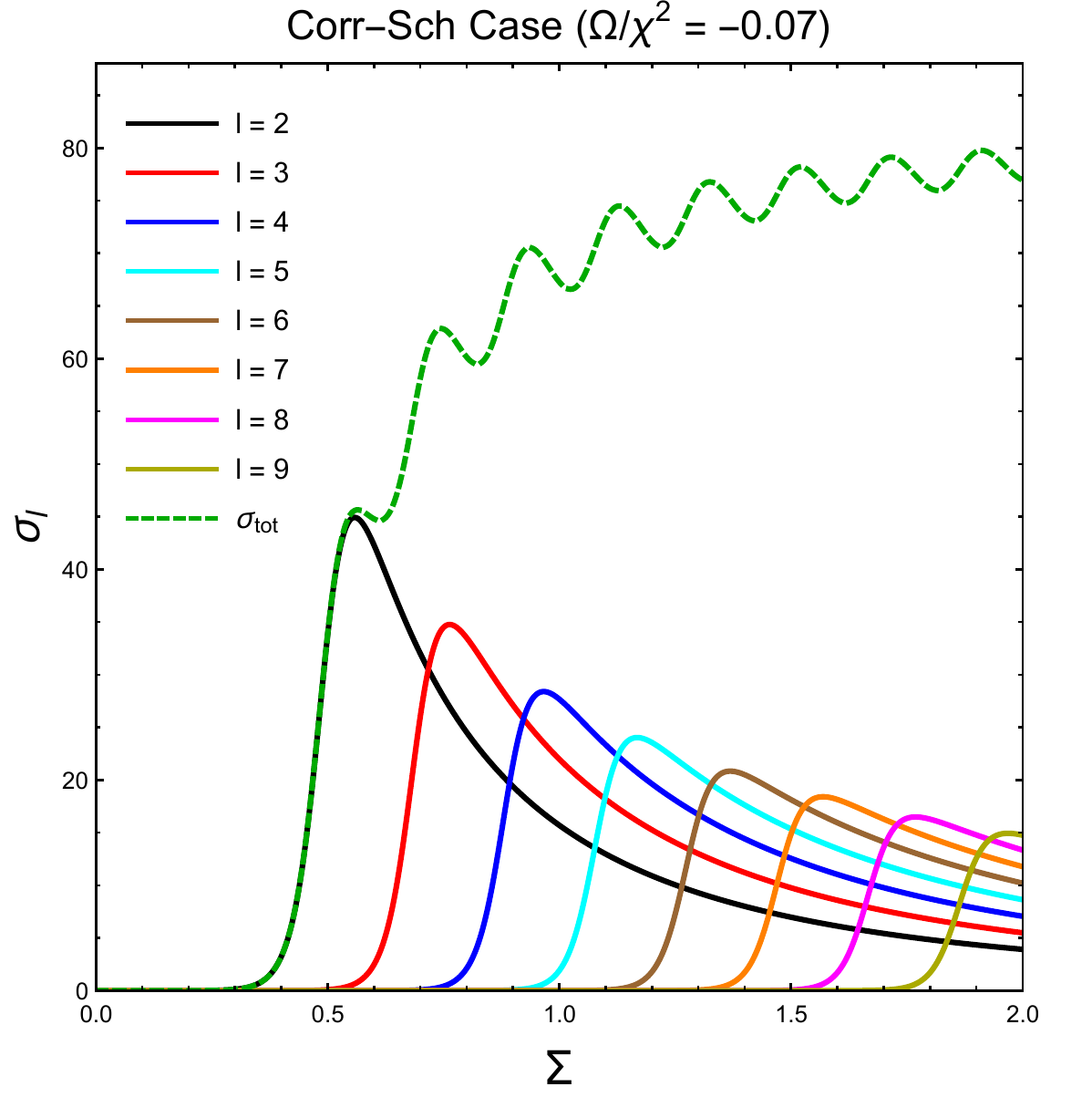}
	\caption{The partial absorption cross section with different value of $l$. The last color corresponds to the total absorption cross section up to $20$ multipole number, $l$.}
	\label{fig:sigma}
\end{figure}

From the information above, we can calculate how far the quantum-motivated BH can absorb the wave. By employing the quantum mechanics analogy, partial absorption cross sections can be written as
\begin{equation}
\sigma_l(\Sigma)= \frac{\pi}{\Sigma^2}(2l+1) ~\Gamma_{l} (\Sigma),
\end{equation}
where the $\Sigma$ is denoted as the continuous frequency, and the $\Gamma_{l}$ is the transmission coefficient $\sim$ $|T|^2$. The partial absorption cross section measures how efficiently the black hole absorbs each $l$ value. Therefore, it is intriguing to observe these properties of the quantum-corrected black hole in the Schwarzschild case. We have plotted the profiles in Fig.~\ref{fig:sigma} for various $l$. At $l=2$, the barrier potential is the smallest compared to other $l$, so the scalar waves can penetrate the barrier more easily and are absorbed by the BH with a high value of $\sigma_{l}$. As $l$ increases, the partial absorption decreases. This is physically expected because higher-$l$ waves can penetrate the barrier and are absorbed less. The total absorption cross section is obtained by summing over all possible $l$. The peak associated with each partial wave occurs at a different frequency since the effective Regge–Wheeler potential depends explicitly on the angular momentum number $l$. As $l$ increases, the centrifugal contribution raises the potential barrier, requiring higher frequencies for significant transmission.

\section{Photon sphere and shadow radius}
\label{Sec: shadow}

In addition to the perturbation, we also study the geodesic properties by considering the motion of the particle by the BH with quantum properties. We particularly focus on the massless particle motion\textemdash specifically photon\textemdash around the BH, and compute the photon sphere and shadow radius.

The motion of a test particle is governed by the geodesic equation,
\begin{equation}
\frac{d\dot{x}^{\mu}}{d\tau} + \Gamma^{\mu}_{\alpha\beta} \dot{x}^{\alpha} \dot{x}^{\beta} = 0,
\label{eq. geodesic equation}
\end{equation}
where $\tau$ is an Affine parameter, $x^{\mu}=(t,r,\theta,\phi)$ and $\dot{x}^{\mu}\equiv dx^{\mu}/d\tau$. 
Restricting motion to the equatorial plane ($\theta = \pi/2,\, \dot{\theta} = 0$), one obtains two conserved quantities, $f(r)\dot{t} = E$ and $\qquad h(r)r^2\dot{\phi} = L$,
which can be identified as the test particle’s conserved energy and angular momentum, respectively. The four-velocity normalization condition for null geodesics is given by
\begin{equation}
g_{\mu\nu}\dot{x}^\mu\dot{x}^\nu=-f(r)\dot{t}^2 + \frac{1}{f(r)}\dot{r}^2 + h(r)r^2\dot{\phi}^2 = 0.
\label{eq. 4-velocity condition}
\end{equation}
Substituting the constants of motion $E$ and $L$ into Eq.~\eqref{eq. 4-velocity condition} yields
\begin{equation}
\left(\frac{dr}{d\tau}\right)^2+V(r)L^2=E^2,\qquad V(r) \equiv \frac{f(r)}{r^2},
\label{eq. geodesic T+V=E}
\end{equation}
where $V(r)$ is known as the photon effective potential~\cite{Claudel:2000yi,Perlick:2004tq}. Applying the chain rule to Eq.~\eqref{eq. geodesic T+V=E}, one have
\begin{equation}
\frac{dr}{d\phi} = h(r)r^2 \sqrt{1/b^2 - V(r)},\qquad  
\label{trajectory1}
\end{equation}
where $b\equiv L/E$ is the impact parameter. 

The photon sphere corresponds to a circular photon orbit at radius $r_{\rm ps}$, obtained by imposing the conditions
\begin{equation}
\frac{dr}{d\phi}\bigg|_{r=r_{\rm ps}}=
\frac{d^2r}{d\phi^2}\bigg|_{r=r_{\rm ps}} = 0,
\end{equation}
which is equivalent to extremizing the effective potential
\begin{equation}
\partial_r V(r_{\rm ps}^\pm) = 0, \label{eq. rps}
\end{equation} 
with stability determined by $\partial_{rr}V(r_{ps}^+)>0$ or $\partial_{rr}V(r_{ps}^-)<0$. Since we have a remarkably simple first order correction to the metric, the photon sphere radius can be obtained analytically, that is
\begin{equation}
r_{\rm ps}^\pm = \frac{1}{4}\left[\sqrt{\eta} + 3\chi \pm \sqrt{-\eta+27\chi^2\left(1+\frac{2\chi}{\sqrt{\eta}}\right)}\right]
\end{equation}
where
\begin{align}
\eta = &4(4\sqrt{3})^{1/3}\gamma + 9\chi^2-\frac{64(6^{1/3})\chi^2\Omega}{\gamma}\\
\gamma=&\left(-27\chi^4\Omega+\sqrt{729\chi^8\Omega^2+2048\sqrt{3}\chi^6\Omega^3}\right)^{1/3}
\end{align}
When $\Omega\to0$, we have $\eta|_{\Omega\to0}=9\chi^2$ and the resulting outer photon sphere $r_{\rm ps}^+$ return to its classical Schwarzschild value of $r_{\rm ps}^+=3\chi$.
Additionally, We will only consider the outer photon sphere $r_{\rm ps}^+$ in the following discussion and drop the $+$ superscript. 

Imposing the $dr/d\phi=0$ condition, one obtain the corresponding critical impact parameter for photon orbit
\begin{equation}
b_c=\sqrt{\frac{1}{V(r_{\rm ps})}}.
\label{eq. crit impact parameter bc}
\end{equation}
Then, the shadow radius $r_{\rm sh}$ observed by a distant observer in an asymptotically flat spacetime can be approximated by a simple relation~\cite{Perlick:2021aok},
\begin{equation}
r_{\rm sh}  \approx b_{\rm c}
\label{eq. shadow rad}
\end{equation}

\begin{figure*}[t!]
	\includegraphics[width=0.46\linewidth]{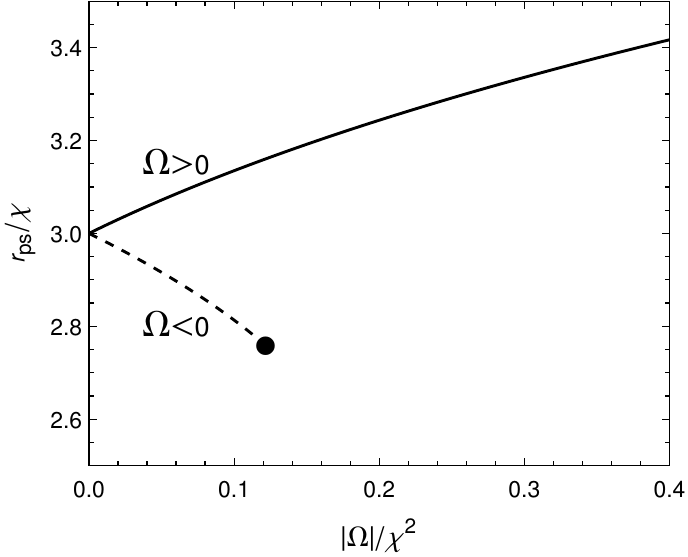}
	\includegraphics[width=0.46\linewidth]{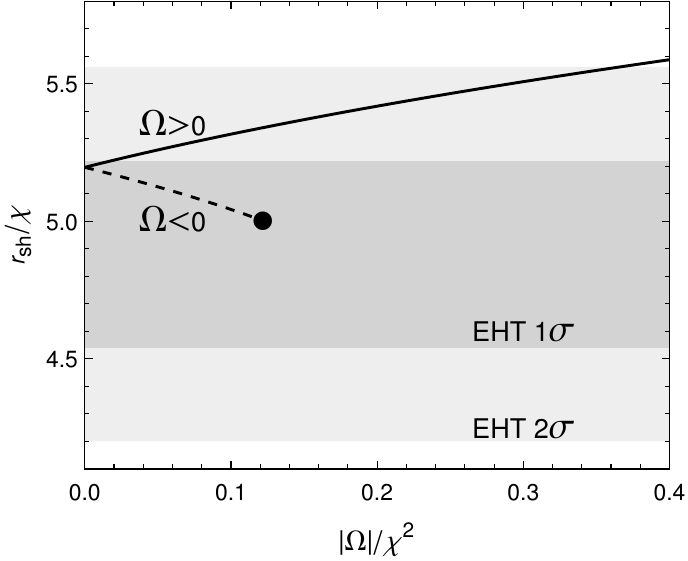}
	\caption{Photon sphere radius $r_{ps}$ (left) and shadow radius $r_{sh}$ (right) as a function of $\Omega$, where the solid and dashed lines represents the case for $\Omega>0$ and $\Omega<0$, respectively. In the $\Omega<0$ case, the curve stops at the extremal BH configuration (c.f. Fig.~\ref{fig:fz_Schwarz}) presented as the filled circle.}
	\label{fig:rshrps}
\end{figure*}

The EHT bound on the shadow radius of the supermassive Sgr A* BH is commonly employed to examine whether the observation results gives a meaningful constraint to the theory (see Refs.~\cite{AlGhifari:2026vil, Bonanno:2025dry, Raza:2025ohk, Jha:2024eyb, Cao:2024vtq, Vachher:2024ait, Ali:2024ssf, Heidari:2024bkm, Chen:2024mlr, Konoplya:2024lch}). From the analysis of Keck and VLTI observational data, Vagnozzi \textit{et al.}~\cite{Vagnozzi:2022moj} derived the following bounds on the dimensionless shadow radius (see also Ref.~\cite{EventHorizonTelescope:2022xqj})

\begin{align}
4.21 < r_{\rm sh}/M < 5.56 \qquad &(2\sigma),\\
4.58 < r_{\rm sh}/M < 5.21 \qquad &(1\sigma).
\end{align}
The resulting shadow radius as a function of $\Omega$ is shown in Fig.~\ref{fig:rshrps}, and the Sgr A* constraints are indicated by the shaded regions. 

Based on Fig.~\ref{fig:rshrps}, the sign of the parameter $\Omega$ fundamentally dictates the scale of both the BH's photon sphere and shadow radius. Positive values of $\Omega$ lead to a clear increase in the sizes of both the photon sphere and the black hole shadow as $|\Omega|/\chi^2$ grows. Conversely, negative $\Omega$ produces the exact opposite trend and results in increasingly smaller radii. Moreover, the parameter space for negative $\Omega$ configurations is physically truncated. The curve stops abruptly at a filled circle representing the extremal BH configuration.

This theoretical truncation carries significant implications for applying observational data because the shadow radius for $\Omega<0$ naturally falls closer to the center of the allowed observational bounds. Consequently, EHT measurements primarily establish definitive upper limits on positive $\Omega$ configurations rather than negative ones. Applying the strict EHT $1\sigma$ upper bound yields a very tight constraint on the model and restricts the highest allowable value of the parameter to $\Omega/\chi^2 \lesssim 0.0105$. However, using the broader $2\sigma$ upper bound allows the model to accommodate a much larger parameter space, with allowable values up to $\Omega/\chi^2 \lesssim 0.3649$.


\section{Conclusions}\label{Sec 5: Conclusions}

We have analyzed an effective theory of quantum Schwarzschild BHs. In this theory, we expand the constant $G_N M_P^2$ in powers of the Kretschmann scalar rather than the proper distance. The first-order metric has a simple closed form. It reduces to the Schwarzschild solution at $\Omega=0$ and has one of three horizon structures depending on $\Omega$:
\begin{itemize}
	\item For $\Omega>0$, there is a single, Schwarzschild-like horizon.
	\item For $\Omega_{1,e}<\Omega<0$, there are inner and outer horizons.
	\item At $\Omega=\Omega_{1,e}=-9\sqrt{3}/128$ (for $\chi=1$), the two horizons merge at $z_e=3/2$.
\end{itemize}
For $\Omega<\Omega_{1,e}$, the solution describes a naked singularity. To second order, $\Omega_2>0$ only shifts the classical lapse function, while $\Omega_2<0$ reproduces the double-horizon structure of the first-order metric. We therefore need only the phenomenological analysis of the first-order metric, which captures the leading infrared (IR) behavior.

For $\Omega<0$, the Hawking temperature vanishes at a finite horizon radius. The entropy and heat capacity then indicate a stable, cold remnant, which is consistent with the third law. For $\Omega\geq 0$, the temperature diverges as $z_+\to 0$. The entropy does not vanish and $C<0$ everywhere, so there is no remnant and no phase transition. Only the branch $\Omega<0$ therefore shows quantum thermodynamic behavior. At large $z_+$, all quantities recover the classical Schwarzschild results.

In the perturbation case, we analyze the first-order Schwarzschild quantum black hole. With the correction determined by $\Omega/\chi^2$, we show that the negative branch transitions from the double horizon to the extreme case. 

In the effective potential, there exist an inner and an outer light ring, while the extreme case shows only the outer one. Based on these results, we use the outer light ring for the subsequent analysis. In quasinormal modes, the frequency is complex: the real part describes the oscillation frequency, and the imaginary part describes the inverse damping time. The black hole with $\Omega/\chi^2 = -0.07$ allows a frequency higher than the others. Consequently, this case decays more slowly. Moreover, we analyze the graybody factor using the WKB method, and transmission exists starting from a certain continuous frequency $\Sigma$. We also explored whether the correspondence between QNM and the graybody still holds by computing the first second-overtone number and the graybody factor. We infer that the higher-order mode closely mimics the graybody profile, with a very small error. Lastly, we investigated the partial and total absorption cross section. We infer that as frequency increases, the cross-section area becomes larger.

On the optical observables side, the parameter $\Omega$ fundamentally governs the characteristic scales of both the BH shadow and the photon sphere. The distinct geometric behaviors driven by the sign of $\Omega$ lead to opposite behavior in the photon sphere and shadow radii. Positive $\Omega$ configurations are directly constrained by EHT shadow measurements, which impose $1\sigma$ and $2\sigma$ upper limits of $\Omega/\chi^2 \lesssim 0.0105$ and $\Omega/\chi^2 \lesssim 0.3649$, respectively.

Overall, the Kretschmann-scalar expansion gives a quantum-corrected Schwarzschild BH. It is distinguishable from GR in the regime $\Omega<0$, and it recovers GR when $\Omega=0$ or at large radius or mass. It only shifts the BH horizon at $\Omega>0$. Several extensions can follow naturally for further research, including studying the ringdown from other perturbations, the dynamics of the remnant, and observational bounds on $\Omega$ from future gravitational-wave and higher-precision horizon-scale measurements. 

\section*{Acknowledgements}

BNJ and AS are supported by Hibah Fundamental Directorate General of Higher Education (DIKTI), PKS-178/UN2.RST/HKP.05.00/2026. MFF is supported by the Indonesia Endowment Fund for Education (LPDP) grant from the Ministry of Finance of the Republic of Indonesia.

\section*{Statement of the Conflict of Interest}

All authors hereby declare that we do not have any conflict of interest.

\section*{Data Availability Statement}

Data sharing is not applicable to this article as no data sets were generated or analyzed during the current study.





\nocite{*}


\end{document}